\documentstyle[11pt,aaspp4]{article}

\begin{document}

\title{Resolving Gamma-Ray Burst~000301C with a Gravitational Microlens}

\author{Peter M. Garnavich} 
\affil{University of Notre Dame, Department of Physics, 225 Nieuwland
Science Hall, Notre Dame, IN 46556}
\affil{\tt e-mail: pgarnavi@nd.edu} 

\author{Abraham Loeb, K. Z. Stanek\altaffilmark{1}}
\affil{Harvard-Smithsonian Center for Astrophysics, 60 Garden St.,
Cambridge, MA 02138} 
\affil{\tt e-mail: aloeb@cfa.harvard.edu,
kstanek@cfa.harvard.edu} 
\altaffiltext{1}{Hubble Fellow}

\begin{abstract}

The afterglow of the Gamma-Ray Burst (GRB) 000301C exhibited
achromatic, short time-scale variability that is difficult to
reconcile with the standard relativistic shock model.  We interpret
the observed light curves as a microlensing event superimposed on
power-law flux decays typical of afterglows. In general, a
relativistic GRB shock appears on the sky as a thin ring expanding at
a superluminal speed. Initially the ring is small relative to its
angular separation from the lens and so its flux is magnified by a
constant factor.  As the ring grows and sweeps across the lens its
magnification reaches a maximum. Subsequently, the flux gradually
recovers its unlensed value.  This behavior involves only three free
parameters in its simplest formulation and was predicted theoretically
by Loeb \& Perna (1998).  Fitting the available $R$-band photometric
data of GRB 000301C to a simple model of the microlensing event and a
broken power-law for the afterglow, we find reasonable values for all
the parameters and a reduced $\chi^2/DOF$ parameter of 1.48 compared
with 2.99 for the broken power-law fit alone. The peak magnification
of $\sim 2$ occurred $3.8\;$days after the burst. The entire
optical-IR data imply a width of the GRB ring of order $10$\% of its
radius, similar to theoretical expectations. The angular resolution
provided by microlensing is better than a micro-arcsecond. We infer a
mass of approximately $0.5\;M_\odot$ for a lens located half way to
the source at $z_{\rm s}=2.04$.  A galaxy 2$^{\prime\prime}$ from
GRB~000301C might be the host of the stellar lens, but current data
provides only an upper-limit on its surface brightness at the GRB
position.

\end{abstract}


\section{INTRODUCTION}

The rapid localization of gamma-ray bursts (GRBs) has brought a new
dimension to GRB research by allowing many events to be followed up at
longer wavelengths. Afterglows of bursts have been detected at X-ray
(Costa et al.~1997), optical (van Paradijs et al. 1997) and radio
(Frail et al.~1997) wavelengths.  Precise positions have allowed
redshifts to be measured for a number of GRBs (Metzger et al.~1997),
providing a definitive proof of their cosmological origin.

The afterglow of a GRB is thought to be synchrotron radiation from a
relativistic shock driven into the circumburst environment (Meszaros
\& Rees 1993, 1997; Paczynski \& Rhoads 1994; Katz 1994; Waxman
1997a).  The light curves and spectral energy distributions are well
fit by power-laws as expected from the shock model. There is recent
evidence that at least some GRB are not spherical explosions.  A
broad-band break in the light curve power-law index was predicted for
shocks produced by collimated jets (Rhoads 1997) and such breaks have
been seen in GRB~990510 (Stanek et al. 1999; Harrison et al. 1999),
GRB~991216 (Halpern et al. 2000), and GRB~000301C (Sagar et al. 2000;
Masetti et al. 2000; Jensen et al. 2000; Berger et al. 2000).  The
ratio of the spectral index to the light curve index also suggests
non-spherical energy ejection for some events (e.g. GRB~991216:
Garnavich et al. 2000a).  In general, the synchrotron afterglow model
has been very successful in matching most of the observations.
However, it is heavily strained explaining the well-studied afterglow
of GRB~000301C which shows a peculiar achromatic fluctuation which
deviates significantly from the broken power-law fit to the lightcurve
(Sagar et al. 2000; Berger et al. 2000).

Here we propose an elegant solution to the GRB~000301C problem: the
GRB was microlensed.  Previously, Loeb \& Perna (1998) predicted that
microlensing by a solar mass lens at a cosmological distance would
produce a nearly achromatic fluctuation of similar amplitude and duration to
that observed.  In \S 2 we describe the optical and near-IR
observations available for our analysis. In \S 3 we describe the
simplest parameterization of a GRB microlensing event and discuss our
technique in fitting the data. In \S 4 we analyze the results and
estimate the probability that a microlensing event can occur given the
observational constraints. Finally, we summarize our conclusions in \S
5.

\section{THE BURST}

GRB~000301C was detected by ASM, Ulysses and NEAR on 2000 March
1.41085 (UT) and localized to a region of about 50 arcmin$^2$ on the
sky (Smith, Hurley \& Cline 2000).  An optical transient was detected
about $1.5\;$days after the burst (Fynbo et al. 2000) and its presence
confirmed in the near-IR (Stecklum et al. 2000) and at radio
wavelengths (Bertoldi 2000). The initial optical decline was steep
with a power-law index of $\alpha=1.6\pm 0.3$ (Halpern, Mirabal \&
Lawrence 2000), but by $3\;$days after the burst the decay rate slowed
down (Garnavich et al. 2000b), a very unusual occurrence in an optical
afterglow. This ``standstill'' did not last long and two weeks after
the burst the power-law index was $\alpha =2.7$ (Veillet
2000b). Rhoads \& Fruchter (2000) found that their near-IR data
behaved differently than the optical although the time sampling was
more sparse in the IR.

Spectra of the afterglow showed a strong Lyman cutoff in the UV
(Smette et al. 2000) and absorption features of Ly$\alpha$ and metal
lines were consistent with a large absorbing column at $z=2.04$
(Jensen et al. 2000; Feng, Wang \& Wheeler 2000).  This could be gas
in the host or an intervening galaxy, but the lack of a Lyman break
longward of 318~nm (Feng et al.~2000) places an upper limit on the
redshift of the GRB of $z_{\rm s}<2.5$.

The most complete photometric compilation is by Sagar et al. (2000)
(shown in their Fig.~3) and is roughly described by two power-laws
with a break around $7\;$days after the burst. But there are clearly
significant deviations from the broken power-law model, especially
between 3 and $6\;$days after the burst.  A similar break was
seen in GRB~990510 (e.g. Stanek et al.~1999) and interpreted as the
lateral spreading of the jet as its Lorentz factor dropped below the
inverse of its opening angle.  But GRB~000301C is unique in showing
variability on the relatively short time-scales of hours to a few days
superimposed on the more typical power-law trends. As noted by
Berger et al.~(2000) and Sagar et al.~(2000), the rapid fluctuation is
achromatic over a range of more than a factor of 5 in wavelength.
Achromatic brightening of a source can be a sign of gravitational
lensing, but the time-scale for microlensing a point source at $z>2$
is substantially longer than a few days due to the low velocities
($\sim 10^{-3}c$) of conventional sources and lenses.  However, Loeb
\& Perna (1998) showed that the superluminal motion of a GRB source on
the sky generically results in microlensing events with durations of
hours to days.

For our analysis, we use the compilation of Sagar et al. (2000) with
additional $UBVI$ photometry (one point in each band) from Stanek et
al. (2000). One $R$-band point has been removed because of its large
deviation from other data taken near the same time. This leaves a
total of 104 photometric data points in seven photometric bands, with
the following distribution of points: $N(U,B,V,R,I,J,K) =
(6,18,8,46,16,3,7)$.

\section{ANALYSIS}

\subsection{Microlensing of a GRB}

A spherical GRB fireball appears on the sky as a thin ring that
exhibits superluminal expansion at a speed $\sim \gamma c$, where
$\gamma$ is the Lorentz factor of the GRB shock (Waxman 1997b).
Figure~3 of Sagar et al. (2000) shows that the achromatic fluctuation
in the lightcurve of GRB~0000301C occurred well before the power-law
break, and so we assume that the during the microlensing event the jet
still behaves as if it is part of a spherical fireball.

We use $t$ to denote the observed time since the GRB trigger in units
of days.  There are two sets of parameters that will define the light
curves.  The first set describes the intrinsic GRB light decays:
power-law slopes, $\alpha_1$ and $\alpha_2$, parameter $\beta$ that
describes the smoothness of the transition between them, and the
transition time $t_b$. Each wavelength band also requires a parameter
that sets the zero point of the power-law.

Loeb \& Perna (1998) showed that the simplest microlensing model can be
described by three parameters (which are all constants), namely: $R_0$, $b$
and $W$. They are defined as follows:

\noindent
{\bf (i)} $R_0$ is the ring radius at $t=1~{\rm day}$ in units of the
Einstein radius of the lens. At other times the ring radius evolves as
(Waxman 1997b)
\begin{equation}
R_s(t)=R_0t^{5/8} .
\end{equation}
Here $R_0={\rho_0/ r_E}$, and
\begin{equation}
\rho_0= 4\times 10^{16}\left({E_{53}\over
n_1}\right)^{1\over 8}(1+z_{\rm s})^{-5/8}~{\rm cm},
\end{equation}
where the factor involving the source redshift $z_{\rm s}$ is due to the
cosmic time dilation, $E_{53}$ is the ``isotropic-equivalent'' of the
energy release in units of $10^{53}~{\rm erg~s^{-1}}$, and $n_1$ is the
ambient gas density in units of ~$1~{\rm cm^{-3}}$. The actual energy
release could, of course, be much smaller due to the small solid angle
occupied by the jet, but this does not affect the source size until the
break in the lightcurve. The Einstein radius of a lens of mass $M_{\rm
lens}$ is,
\begin{equation}
r_E=\left[\left({4GM_{\rm lens}\over c^2}\right)D\right]^{1/2}=
7.7\times 10^{16} \left({M_{\rm lens}\over 1
M_\odot}\right)^{1/2}\left({D\over 10^{28}~{\rm cm}}\right)^{1/2}~{\rm cm}
\end{equation}
where $D\equiv ({D_{\rm s}D_{\rm ls}/ D_{\rm l}})$ is the ratio of the
angular-diameter distances between the observer and the source, the lens
and the source, and the lens and the observer. The value of $D$ depends on
the lens and source redshifts and the cosmological parameters. 

\noindent
{\bf{(ii)}} $b$ is the lens-source separation on the sky (or equivalently,
the ``impact parameter'') in units of the Einstein angle, $\theta_E\equiv
(r_E/D_{\rm s})$.  Since the apparent source radius is changing faster than
the speed of light, all other non-relativistic astrophysical motions are
irrelevant. We may therefore consider the source-center and lens positions
as fixed on the sky. Initially, as long as $R_{\rm s}\ll b$ the source is
pointlike and is magnified by a time-independent factor.  The maximum in
the magnification curve is reached when $R_{\rm s}\approx b$; the timing of
this maximum can be used to fix the ratio $R_0/b$.

\noindent
{\bf (iii)} $W$ is the brightness-weighted width of the ring divided by its
radius. The width determines the height and the duration of the
magnification event. The larger the width, the smaller the height is and
the broader the duration is. This width was estimated theoretically to be
$\sim 10\%$ in the optical-IR (see, e.g. Waxman 1997b; Sari 1998;
Panaitescu \& Meszaros 1998; Granot et al. 1999) but is still subject to
uncertainties concerning the extent of the emission region behind the GRB
shock front.  Note that the assumption that the ring has sharp boundaries
and that $R_0$ and $W$ are the same for all wavelengths, is an
over-simplification (see ring profiles in Granot et al. 1999).
Ring properties that vary significantly with wavelength would produce different
light curves in the various observed bands (Wambsganss \& Paczy\'nski 1991). But,
more parameters are required to fit a more elaborate model and this is not
justified by the quality of the available data on GRB~000301C.

If $W$ were close to unity, then the lensing signal would be very
weak. Thus, GRB~000301C provides the first evidence that a GRB produces a
ring on the sky. Previous data, such as the scintillations of GRB~970508,
constrained the source size but not its shape (Waxman, Kulkarni, \& Frail
1998).

\subsection{Magnification Factor}

The magnification factor $\mu$ is a function of $R_s$, $W$, and $b$.  As
described by Loeb \& Perna (1998),
\begin{equation}
\mu(R_s,W,b)={{\Psi}[R_s,b]-(1-W)^2{\Psi}[(1-W)R_s,b]\over 1 -
(1-W)^2},
\label{eq:mu}
\end{equation}
where ${\Psi}(R_s,b)$ is the magnification for a uniform disk of radius
$R_s$ (Schneider, Falco, \& Ehlers 1992),
\begin{equation}
{\Psi}[R_s,b]= \frac{2}{\pi R_s^2}\left[\int_{|b-R_s|}^{b+R_s} dr
\frac{r^2+2}{\sqrt{r^2+4}}\arccos\frac{b^2+r^2-R_s^2}{2rb} +
H(R_s-b)\frac{\pi}{2}(R_s-b)\sqrt{(R_s-b)^2+4}\right]\;.
\label{eq:circ}
\end{equation}
Here $H(x)$ is the Heaviside step function.  The integral in equation
(\ref{eq:circ}) can be expressed more explicitly as a sum of elliptic
integrals (Witt \& Mao 1994). In general, an arbitrary ring profile can
be incorporated as a sum over a set of infinitesimal rings.

\subsection{Fitting the Data}

Using all the data for GRB~000301C (104 points), we performed $\chi^2$
minimization fits using broken power-law model (11 free parameters: 4 shape
parameters $+$ 7 photometric zero points), in the form described by Sagar
et al.~(2000), and using broken power-law plus microlensing model (14 free
parameters: 11 parameters $+$ 3 microlensing parameters).  We repeated the
calculation using only the $R$-band data (5 and 8 free parameters,
respectively), since it had the most complete temporal coverage with the
total of 46 points.

For the broken power-law model, the fits were clearly not good, with
$\chi^2/DOF=2.76$ for all seven bands and $\chi^2/DOF=2.99$ for the
$R$-band data only, values similar to those found in the other papers
describing GRB~000301C (e.g. Jensen et al.~2000). Adding the microlensing
magnification has substantially improved the fits, with $\chi^2/DOF=1.77$
for all seven bands and $\chi^2/DOF=1.48$ for the $R$-band data only.

\begin{figure}[p]
\plotfiddle{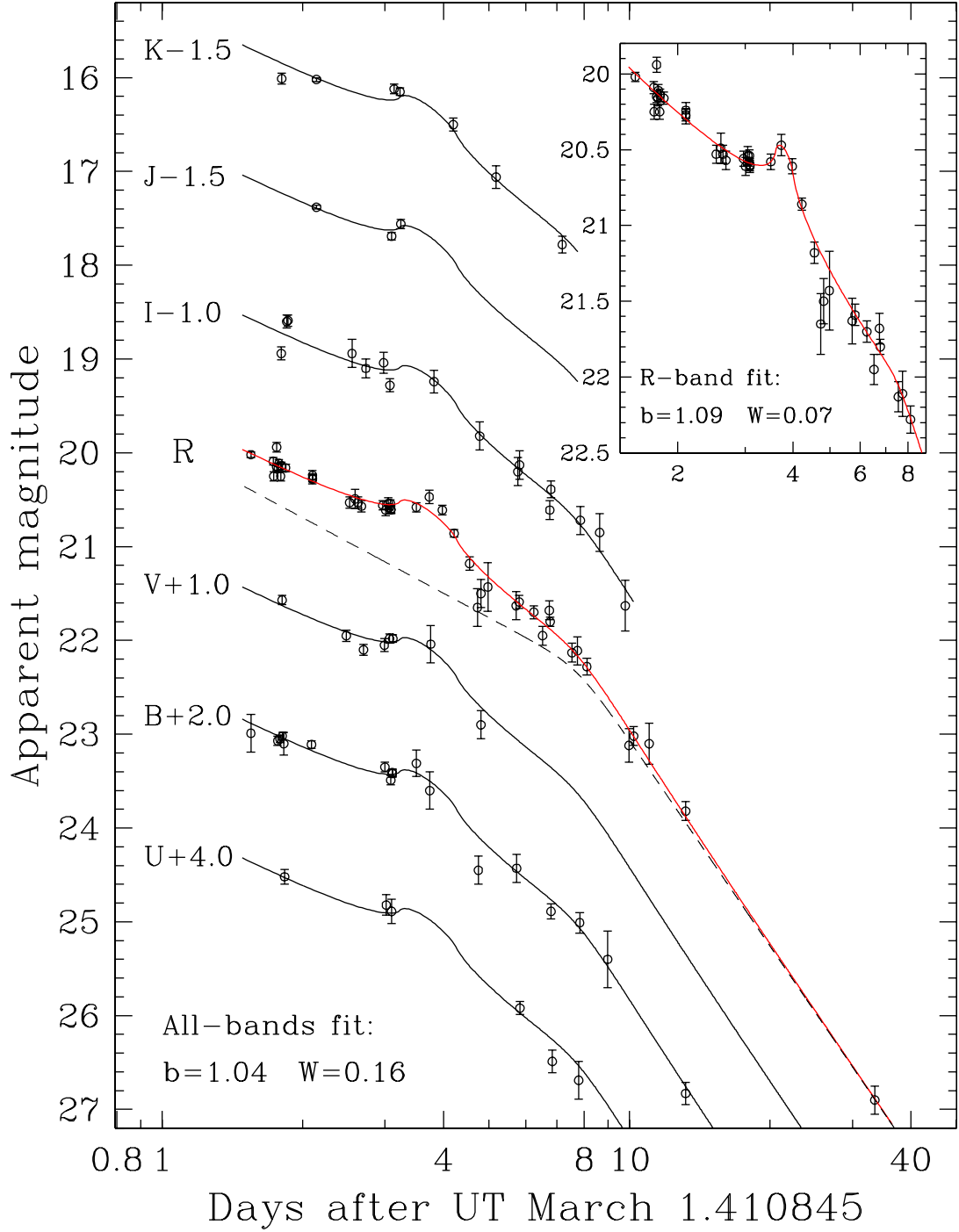}{18cm}{0}{75}{75}{-240}{-30}
\caption{The $UBVRIJK$ light curves of GRB~000301C from the Sagar et
al.~(2000) compilation plus data from Stanek et al.~(2000).  The solid
line is the best fit broken power-law$+$microlensing model obtained
using all the data. The dashed line shows the expected GRB light curve
if it had not been lensed. The insert shows the best fit model using
the $R$-band points only.}
\label{fig:micro}
\end{figure}

\begin{figure}[t]
\plotfiddle{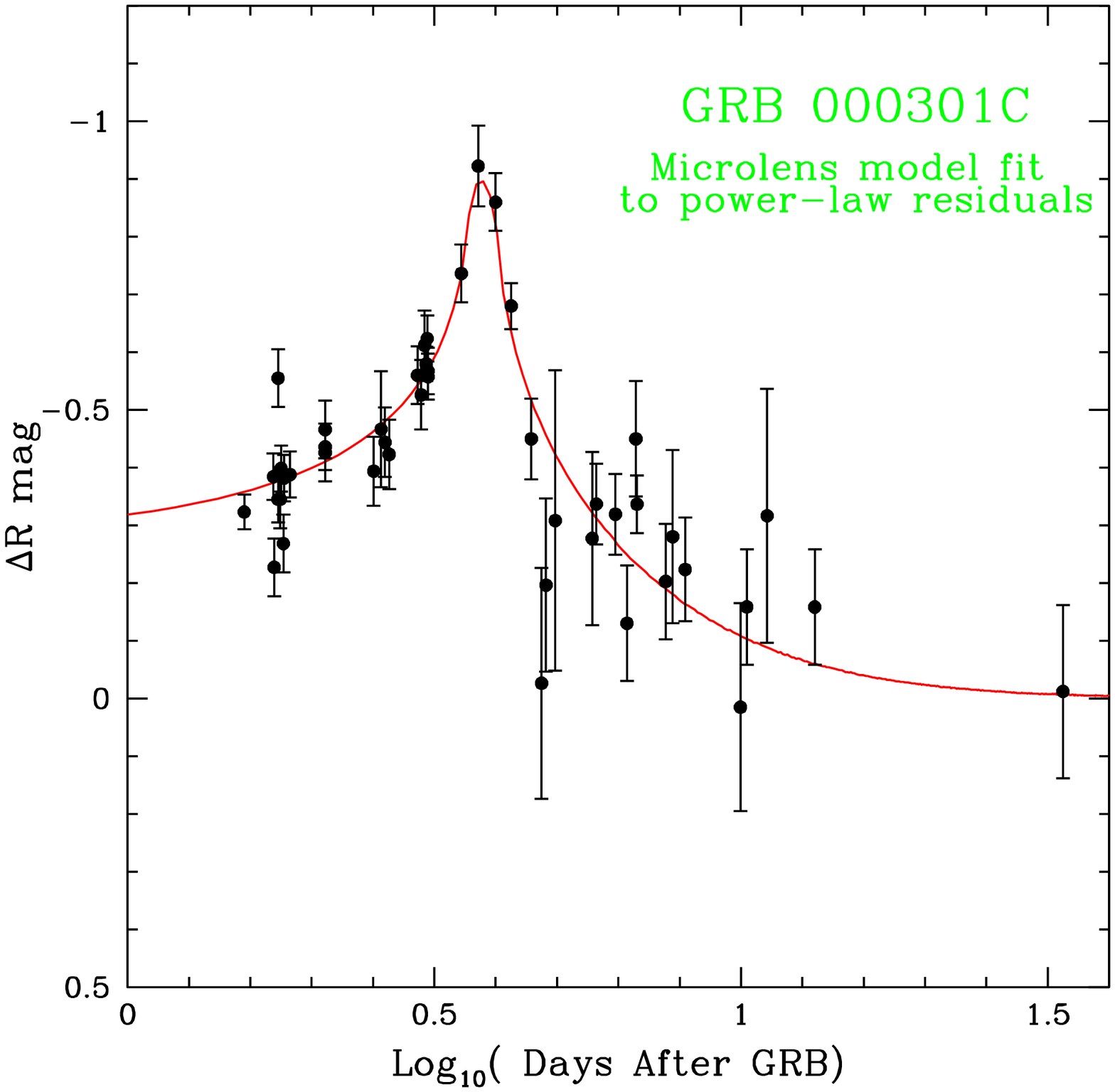}{9cm}{0}{55}{55}{-170}{-90}
\caption{The $R$-band light curve of GRB~000301C with the
best fit broken power-law model removed. The solid line
shows the microlensing model. }
\label{fig:resid}
\end{figure}

The best-fit microlensing models are shown in Figure~\ref{fig:micro}
for all seven bands. Also shown, in an insert, is the best-fit model
for the $R$-band data only.  The best-fit to all data provides an
impact parameter $b=1.04 \pm 0.02$ and a fractional ring width of
$W=0.16 \pm 0.02$. The best-fit to the $R$-band data only yields an
impact parameter $b=1.09 \pm 0.02$ and a substantially smaller width
of the ring, $W=0.07 \pm 0.02$.  In both cases $R_0= 0.49\pm0.02$, and
microlensing provides an angular resolution better than a
micro-arcsecond in probing the GRB fireball. The values for the broken
power-law parameters obtained by us are very similar to those obtained
by Sagar et al.~(2000): $\alpha_1=1.1, \alpha_2=2.9$ and $t_b=7.6\;
$days, regardless if all the data or $R$-band only were fitted. The
value of $\beta$ is not well constrained, but it is always large,
$\beta>5$, indicating a sharp break. The effect of lensing is shown in
Figure~2 after our best fit broken power-law is removed.

The errors in the microlensing parameters estimations are based on
conditional probability distributions, obtained by fixing the rest of
the parameters at their most probable values, and should be treated
only as rough estimates.  It is clear from Figure~\ref{fig:micro} that
the actual errors of the photometry are not Gaussian, but are
dominated by systematic errors, probably resulting from different
reduction procedures applied to the various data.  Given the special
nature of GRB~000301C, it would be well-worth the effort for one of
the groups, possibly one which obtained much of the photometric data
for this afterglow, to reduce the CCD data obtained by the other
groups in order to ensure a uniform reduction procedure.  Such uniform
reduction would allow for better statistical testing of the
GRB~000301C microlensing hypothesis.

\section{DISCUSSION}

The global fit to the data matches the light curves of the individual
bands well. The fit to the $R$-band alone is also of high quality and
gives similar values for the microlensing parameters.  The main
difference between the global and $R$-band is in the sharpness of the
microlensing peak which is narrower and therefore favors smaller
values of $W$ in the $R$-band.

We can estimate the mass of the lens from our measurement of the ring
radius parameter, $R_0$.  The lens mass can be written as
\begin{equation}
M_{\rm lens}=0.13\;M_\odot\times \left({E_{53}\over n_1}\right)^{1/4}
\left({D_{\rm l}\over D_{\rm ls}}\right)\; R_0^{-2}\;\; ,
\end{equation}
assuming a source redshift of $z_{\rm s}=2.04$ and a flat universe with
$h=0.7$ and $\Omega_m=0.3$.  Hence, for $R_0=0.5$, the lens mass is of
order $\sim 0.5\;M_\odot$ i.e. stellar size, for a wide range of reasonable
GRB energies and circumburst densities. This is reassuring for the
microlensing interpretation since it does not rely on some peculiar and
rare object to do the lensing. In a case analogous to GRB sources, Koopmans
et al. (2000) have recently interpreted variability of the macro-lensed
superluminal radio source B1600+434 as being due to microlensing.

\subsection{Probability for Microlensing}

\begin{figure}[t]
\plotfiddle{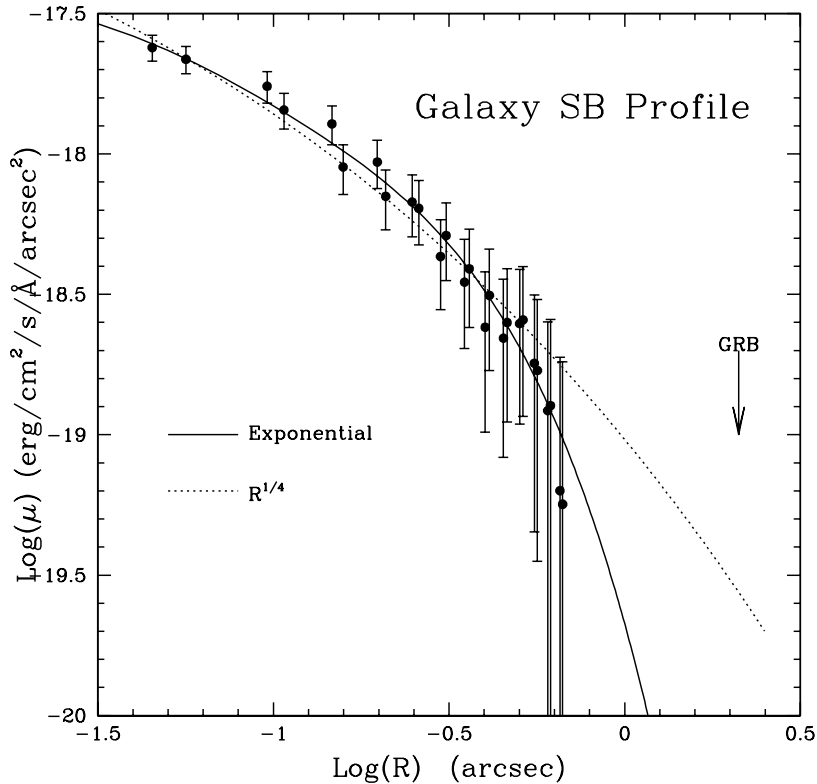}{9cm}{0}{55}{55}{-180}{-90}
\caption{The $V$-band surface brightness distribution along the major axis
of the galaxy 2$''$ from the GRB. The data are from $HST/STIS$ images. The
solid line is a fit using an exponential with a scale length of
0.22$^{\prime\prime}$ plus a central point source. The dotted line is a fit
using an $R^{1/4}$ law. The exponential provides a better fit.  }
\label{fig:galaxy}
\end{figure}

Deep Hubble Space Telescope ({\it HST}) imaging of the GRB and field
were obtained by Fruchter, Metzger \& Petro (2000) but no host or
intervening galaxy was detected to a limit of 28.5 mag. A galaxy with
$R=24.3\pm 0.3$ mag (Veillet 2000a) is 2$''$ from the position of the
GRB. Our analysis of the deep STIS exposures obtained 2000, April 19,
is shown in Figure~\ref{fig:galaxy} and suggests that the surface
brightness profile along the major axis of the galaxy is best fit by
an exponential with a scale length of $0.22''$.  This implies that the
light is from a rather compact disk. The redshift of this nearby
galaxy is unknown.  The galaxy is not detectable out to the position
of the GRB and extrapolation of the surface brightness profile implies
a very low surface brightness of $< 10^{-23}$~${\rm erg\; cm^{-2}\;
s^{-1}\; \AA^{-1}\; arcsec^{-2}}$ at an effective wavelength of
585~nm.  Of course, a dark halo may extend well out beyond the visible
disk.  From the background variations in the $HST$ images, we place a
limit on the surface brightness at the GRB of $> 28.5 $~${\rm mag\;
arcsec^{-2}}$ in approximately the $V$-band.

{\it What is the probability for microlensing by a solar mass star at a
cosmological distance?} This probability can be written as $P_{\rm
ml}=\Sigma_\star/\Sigma_{\rm crit}$, where $\Sigma_\star$ is the surface
mass density of stars in an intervening galaxy close to the line-of-sight
and $\Sigma_{\rm crit}=(c^2/4\pi G)(D_{\rm s}/D_{\rm l}D_{\rm
ls})b^{-2}$. If the mass-to-light ratio of these stars is $(M/L)$, then
their apparent surface brightness is $\mu_{\rm
SB}=(M/L)^{-1}(\Sigma_\star/4\pi(1+z_{\rm l})^4)$. Hence the expected
surface brightness of stars around the location of the afterglow is given
by,
\begin{equation} 
\mu_{\rm SB}=\left({c^2\over 16\pi^2G b^2}\right)\left({M\over
L}\right)^{-1} \left({D_{\rm s}\over D_{\rm l}D_{\rm ls}}\right) {P_{\rm
ml}\over (1+z_{\rm l})^4} .
\end{equation} 
The simple point-lens model is adequate for $P_{\rm ml}\ll 1$, so that
the caustics induced by the external shear occupy a region much
smaller than $r_E$ (Chang \& Refsdal 1984).  For $P_{\rm ml}\sim 0.1$,
$z_{\rm s}=2$, $b=1$, $(M/L)=5$ in solar units, and a flat universe
with $h=0.7$ and $\Omega_m=0.3$, we find $\mu_{SB}\approx 29$
magnitudes per arcsec$^2$ in the V-band, for a lens galaxy at $1\la
z_{\rm l}\la 1.7$, assuming a $K$-correction of $\sim 4$ V-magnitudes
as required for an elliptical galaxy in this redshift interval (see
Fig. 10 of Fukugita et al. 1995). This value of $\mu_{SB}$ is below
the inferred upper limit on the surface brightness from the {\rm HST}
image.  Coincidentally, it is comparable to the $R^{1/4}$--law
extrapolation of the surface brightness in Figure 3, but well above
the exponential extrapolation.  The above constraint is much weaker if
the lensing star belongs to a halo population of compact objects
(MACHOs) which have a high $M/L$ ratio.  The microlensing probability
would inevitably be large if the HI column density $\sim 10^{21}~{\rm
cm^{-2}}$, detected at $z=2$ (Jensen et al. 2000; Smette et al. 2000),
is due to the galactic host of the lens (Perna \& Loeb 1997); however
the observed damped Ly$\alpha$ absorber is more likely to be the host
galaxy of the GRB.

\subsection{Alternative Interpretations}

Fluctuations in the lightcurve may result from shell collisions within
the fireball (Dai \& Lu 2000), but such collisions would re-energize
the fireball and change its spectrum by increasing the peak frequency
of its emission. There is no evidence for a {\it chromatic} change of
this type during the event (Berger et al 2000).  The activity of the
central engine of the GRB has to be fine-tuned in an ad-hoc manner so
as to produce a collision only after $\sim 4\;$days when the Lorentz
factor of the expanding shell ($\gamma\sim 5$) already declined by
more than an order of magnitude relative to its initial value.

Berger et al. (2000) suggested an alternative interpretation of the
{\it achromatic} fluctuation in terms of an inhomogeneity of the
ambient density into which the fireball is propagating, $n$.  Naively,
one may argue that the flux from a fireball propagating into a uniform
ambient medium scales as $\propto n^{1/2}$ at all wavelengths, and so
a brightening of the flux by a factor of 1.7 may be achieved if the
fireball encounters a clump of gas which is a factor of $\sim 3$
denser than the mean. The required clump location ($\sim 5\times
10^{17}~{\rm cm}$), transverse size ($\ga 10^{17}~{\rm cm}$), depth
($\la 5\times 10^{17}~{\rm cm}$) and overdensity ($\sim 3$), can all
be chosen ad-hoc so as to fit the data.  However, the standard model
for the fireball dynamics and emission implies that the flux at a
particular observed time is emitted with different weights on a
particular spatial region for frequencies above and below the peak
spectral frequency [see Eqs. (15) and (16) in Wang \& Loeb 1999]. This
results in two effects: (i) the ring is narrower and has a higher
contrast (center to limb variation in brightness) at higher
frequencies [see Figs. (11) and (12) in Granot et al. 1999]; (ii) the
limb of the ring is influenced by the clump region at a later observed
time than its center because of the geometric time delay.  Since the
peak frequency is $\sim 300$GHz at $\sim 4\;$days (Berger et al. 2000;
Sagar et al. 2000), the fluctuation is therefore expected to brighten
to a maximum earlier in the radio than in the optical-IR. The
chromaticity should be even more pronounced below the synchrotron
self-absorption frequency (Berger et al. 2000), as this frequency
depends on $n$.  Although it may not be possible to rule out the above
chromatic behavior with the sparse radio data for GRB 000301C,
achromaticity does not trivially follow from this model at wavelengths
ranging all the way from the optical to the radio.

The ring parameters $W$ and $R_0$ which define the microlensing
lightcurves should also depend on wavelength (Panaitescu \& Meszaros
1998; Granot et al. 1999) but in a generic predictable way. Hence,
with better analysis of the existing data for GRB~000301C or with
detailed monitoring of future GRBs it should be possible to test it
against the above interpretations.

\section{CONCLUSIONS}

We successfully model the light curve of GRB~000301C as a standard
broken power-law plus a gravitational microlensing event. We find the
lens mass required for the event is $\sim 0.5$M$_\odot$, thus
requiring only a normal star to explain the lensing. No galaxy or
diffuse light from the stellar population is detected along the line
of sight, but reasonable lensing probability is expected even for
surface brightness below the deep $HST$ images we analyzed.  Ground
based images with a large telescope may detect the lensing population.

New GRB satellites such as $HETE-2$ and $SWIFT$ will provide many
opportunities to study GRB afterglows. This is the first of what will
be a number of microlensed GRBs and we show that analysis of
high-quality data covering a large wavelength range can be a useful
tool for probing the physical structure of GRB afterglows with sub
micro-arcsecond resolution.  Our estimate of $W$ confirms the
prediction that afterglow shocks appear as thin rings on the sky
(Waxman 1997b; Sari 1998; Panaitescu \& Meszaros 1998; Granot et
al. 1999). Better broad-band data could check for variation with
wavelength of the ring width and radius, and polarimetric observations
could confirm the predictions of the polarization variations made by
Loeb \& Perna (1998). Future observations could also constrain the
properties of the dark matter by taking a census of the number and
masses of microlenses.

\acknowledgments{We thank Bohdan Paczy\'nski and Eli Waxman for useful
comments on the manuscript and Dave Bennett for helpful discussions.
Support for KZS was provided by NASA
through Hubble Fellowship grant HF-01124.01-A from the Space Telescope
Science Institute, which is operated by the Association of
Universities for Research in Astronomy, Inc., under NASA contract
NAS5-26555. This work was supported in part by NSF grant AST-9900877,
the Israel-US BSF, and the NASA grant NAG5-7039 for AL. Support for
PMG was provided by NASA LTSA grant NAG5-9364.}

\end{document}